\documentclass[aps,twocolumn,floatfix,showpacs]{revtex4}
\usepackage{epsfig}
\usepackage{graphicx}
\usepackage{dcolumn}
\usepackage{natbib}
\usepackage{amssymb}
\usepackage{placeins}

\begin{document}

\title{Damage accumulation in quasi-brittle fracture}

\author{Claudio Manzato}
\author{Mikko J. Alava}
\affiliation{COMP Centre of Excellence, Department of Applied Physics, Aalto University,
P.O. Box 14100, FIN-00076, Aalto, Espoo, Finland}
\author{Stefano Zapperi}
 \affiliation{CNR - Consiglio Nazionale delle Ricerche, IENI, Via R. Cozzi 53, 20125,
Milano, Italy}
\affiliation{ISI Foundation, Via Alassio 11/c 10126 Torino, Italy}
\begin{abstract}
The strength of quasi-brittle materials depends on the ensemble of
defects inside the sample and on the way damage accumulates before
failure. Using large scale numerical simulations of the random fuse
model, we investigate the evolution of the microcrack distribution
that is directly related to the strength distribution and its size
effects. We show that the broadening of the distribution tail
originates from the dominating microcracks in each sample and is
related to a tendency of crack coalescence that increases with system
size. We study how the observed behavior depends on the disorder
present in the sample.
\end{abstract}



\maketitle

\section{Introduction}

\begin{figure*}[!ht]
\includegraphics{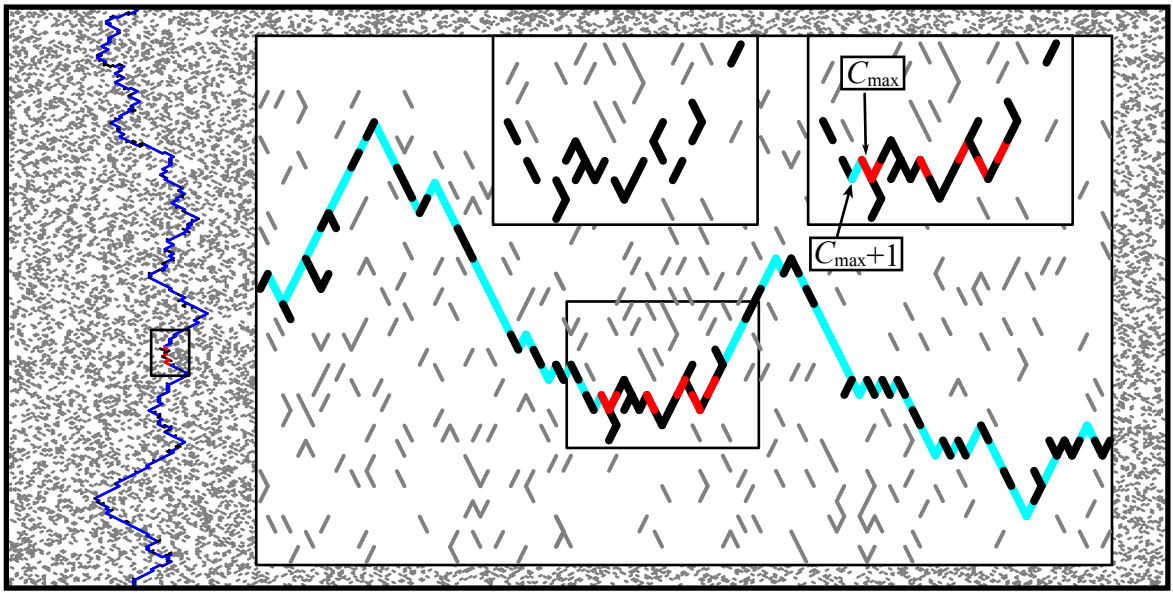}
\caption{(Color online) Sample snapshopt from a $512 \times 512$ system. A fraction
  $n_r=0.10$ of bonds has been removed at the beginning (gray/black,
  black belonging to the final spanning crack). In red the bonds
  broken during the pre-peak damage and, in cyan the ones that belong
  to the final avalanche. Inset: a section of neighborhood for the
  final crack containing the \emph{red} bonds in particular. More
  detail is shown in the two smaller insets: the initial configuration
  (left) and the configuration at the peak (right). The bonds broken
  from the beginning contribute by bridging to increase the length of
  the cracks that will fail the entire sample. $C$ indicates here the
  current, indexed by the number of bonds failed. }
\label{fig:1}
\end{figure*}

The fracture strength of materials is a problem whose general
understanding is still based on empirical knowledge that would benefit
from a more solid fundamental theory (for a review see
\cite{Alava_et:2006, Alava_et:2009}). The simplest  case to study is
the one where there is no time-dependent rheology or memory effects
like in plastic deformation: the only trace that a sample maintains of
its past history is the additional damage that it has accrued during
earlier loading. Then, the most important problem of the fracture
(peak) strength and its size-scaling becomes an exercise in extremal
statistics or the renormalization of distributions of extreme values 
from sub-systems \cite{Manzato_et:2012, Shekhawat_et:2013}. The main
issue can be summarized as follow: does one of the limiting extreme
value distributions (i.e. Gumbel, Weibull, Frechet) \cite{Gumbel:1958}
describe fracture and why? The answer and its explanation  depends, in
the limit of dilute disorder, on the distribution of (micro)cracks
present in the sample. One sub-volume has the largest defect, inducing
failure at the stress that this creates by linear elastic fracture
mechanics in its neighborhood. Engineers have for decades used the
Weibull distribution \cite{Weibull:1939} while in other cases, where
the stress enhancements turn out to be negligible, even Gaussian
(normal) distributions arise (e.g. in the fiber bundle model
\cite{Pradhan_et:2010}).

For initially random disorder, the damage present without any previous
loading implies an exponential defect distribution at least at weak
disorder. If the subsequent damage from loading is neglected, the
problem is directly solvable \cite{Duxbury_et:1986,Duxbury_et:1987} in
terms of a modified Gumbel distribution, which can be shown to flow
asymptotically to the limiting Gumbel case
\cite{Manzato_et:2012}. Damage accumulation does not change this, nor
do eventual interactions via long-range stresses do so
\cite{Shekhawat_et:2013}. However, high-accuracy studies of the defect
distributions at peak-load imply that the very small amount of
additional damage does affect the tails, changing the original
exponential one to a wider one \cite{Manzato_et:2012}.

Here, we analyze this mechanism in detail by considering the
development of the microcracks for various disorders and as a function
of sample size. In the random fuse model simulations we perform to
this end - a discretized scalar fracture model - one can keep track of
the additional element failures, and analyze in detail the microcrack
geometries and densities. The main result we obtain is that the
development of the wider tail is a rare-event phenomenon: in each
sample, the largest crack contributing to that tail is an unique case,
and it typically arises from the coalescence of two smaller
microcracks. This phenomenon exhibits scaling with disorder and sample
size, and indicates how the details of the damage mechanics will then
influence the quantitative stress scaling. This is so since the
general form of the defect population merely dictates the form of the
extremal distribution and scaling of failure stress with sample size,
leaving room for the microscopic detail. 

The structure of the rest of this paper is as follows: first we
present briefly the numerical model, concentrating more on the
analysis of damage and crack evolution. Section 3 presents the
results. First we discuss the development of damage as a function of
various parameters upto the maximum stress (current) and connect it to
the known scaling of the peak stress. Then, we analyze the microcrack
distributions at maximum, and pay particular attention to where the
largest cracks (in each sample) come from. We present a scaling
analysis of their properties with sample size and disorder. Finally,
we look at even more detail at the creation of the largest ones: what
is the role of damage in crack-coalescence.

\section{The model}

We perform numerical simulations of the two dimensional random fuse
model (RFM) \cite{deArcangelis_et:1985}. The model is based on
removing one by one conducting bonds, with unit conductivity
$\sigma_0=1$, from a two-dimensional diamond lattice of size $L\times
L$ \cite{Alava_et:2006}, with periodic boundary conditions in the
direction perpendicular to the application of loading. Initially, a
fraction $n_r$ of the bonds are removed to result in a statistically
homogeneous damage field. In the dilute limit, in particular, it is
easy to see \cite{Duxbury_et:1986} that an exponential initial
microcrack distribution $P_{0,L} (l)$ for defects of size $l$ ensues,
at zero damage ($N=0$).

To analyze the development of the microcrack distribution
$P_{N,L}(l)$, where $N$ now refers to the number of failed bonds/fuses
in each sample, we follow the cracks that are present in each sample
starting from the original one, $P_{0,L} (l)$. Figure \ref{fig:1}
shows an initial damage state and the system at peak load, with
$N_C=5$. We notice the following details in the damage development:
the pre-peak damage is small ($N$ upto $C_{\mathrm{max}}$). Some of that
damage, if not all of it, is concentrated locally around some
pre-existing damage, creating (right small Inset in the Figure) a
micro-crack that turns out to be the critical one. This crack is
created by the coalesence of damage by "bridging bonds". Finally, an
unstable avalanche creates the final fracture surface, which is
indicated by the blue bonds in the Figure.

\section{Damage accumulation and cracks}
The following analyses are conditioned by the percolation threshold
($n_{r,c}=0.5$ for the current geometry). One expects to see different
behavior for the damage development in both the weak disorder and
percolation limits. Larger values of $n_r$ make the study of crack
geometries cumbersome, thus most of the results are confined to $n_r
\leq 0.35$. This also excludes the cross-overs from
percolation-dominated to "bulk-like" fracture that happen with
increasing $L$ in the proximity of the percolation value $n_{r,c}$. We
first look in detail at the relevant damage, before concentrating on
its role in the formation of the critical microcracks and the crack
population dynamics. 

\subsection{Scaling of damage}
\begin{figure}[!hbt]
\includegraphics{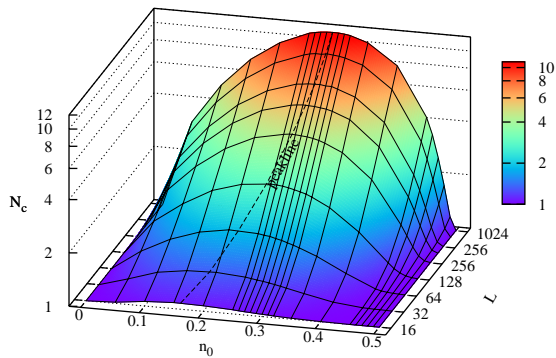}
\caption{(Color online) The figure shows how the amount of pre-peak damage (number of
  broken bonds at maximum current) varies as a function of both the
  disorder $n_0$ and the size of the system $L$ ($N_C$ and $L$ on
  logarithmic scales). For every size a peak develops for some
  $n_0$, which location (dashed line) increase with both the size and
  the initial disorder, but that seems to reach an asymptotic value
  for large sizes.}
\label{fig:2}
\end{figure}
\begin{figure}[!hb]
\includegraphics{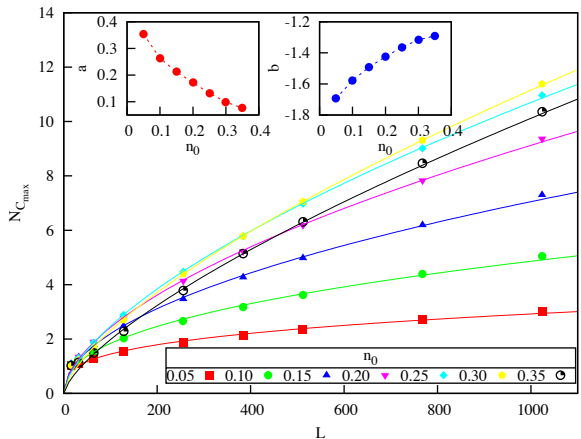}
\caption{(Color online) This figure details the pre-peak damage
  (cfr. Figure~\ref{fig:2}) $N_C$ as a function of the size $L$ for
  various disorders $n_0$. For every fixed disorde $n_0$, $N_C(L)$ is
  well described by a power law $N_C(L)/L^2 = aL^b$. The fitting
  parameters for different disorders are showed in the insets
  (prefactor: left, exponent: right).}
\label{fig:3}
\end{figure}

Figure~\ref{fig:2} shows the pre-peak damage/sample-averaged $N_C$ in
a plot where $L$ and $N_C$ are depicted on logarithmic scales. Note
that the z-axis starts from unity, since one needs always to break at
least one fuse to bring the system to failure. Across the variety of
disorder strengths $n_r$ and $L$-values present here it is clear that
the typical damage is small --- while the strength distribution itself
is of the modified Gumbel-type. In both limits of $n_r$ we see that
the damage approaches the minimum value. In-between, there is a peak
in the damage for a $n_r$-value, which shifts with $L$ and might seem
to saturate (i.e. not approach $n_{r,c}$ as would be the opposite case). 

A more detailed look at the damage scaling in Figure \ref{fig:3}
indicates actually the same peak-effect, as the largest disorder case
($n_r=0.35$) in the Figure illustrates. All the finite size effects
seem to adhere to a power-law increase of $N_C(L)$, which is
sub-extensive, $N_C/L^2 \propto L^b$, with $b$ negative. Note that the
exponent $b$ changes monotonically with $n_r$, and the peak damage
with $n_r$ arises thus from a decrease in the prefactor of the
power-law. This would imply that the apparent asymptotic saturation of
the peak damage in Fig. \ref{fig:2} is just illusory.

\subsection{Crack populations at peak strength}

\begin{figure}[!htb]
\includegraphics{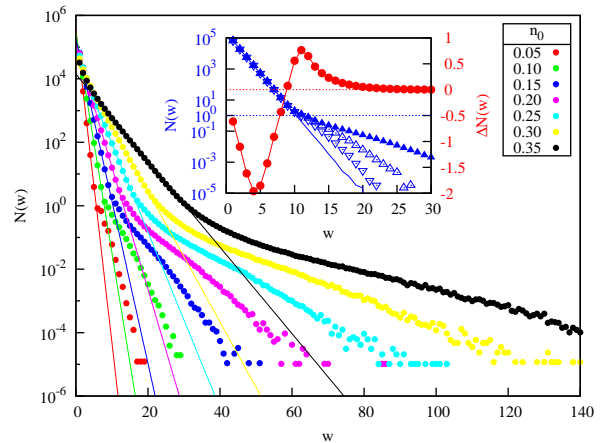}
\caption{(Color online) Average number of clusters $N(w)$ of a given width $w$ per
  sample. The distribution at the beginning is exponential (continuous
  line), while at the peak it develops a tail with a different slope 
  (data points only). The inset shows (in blue) the two distributions
  for a given disorder ($n_0=0.15$) and size ($L=1024$) and (in red)
  their difference. The open-triangle data points present the two
  distributions at the peak obtained subtracting the first- and
  second-largest crack.}
\label{fig:4}
\end{figure}

The main question is now, how does the damage influence the cracks at
peak strength and thus the important macroscopic scaling of sample
strength? Figure \ref{fig:4} shows both a typical result and the main
concepts of a detailed analysis to this end. One can distinguish at
the peak $C$ among several microcrack populations: the original at
zero damage ($N=0$), the one at peak strength, and the one obtained by
subtracting for each sample the first- and second- largest microcracks.

As was already pointed out in Ref. \cite{Manzato_et:2012} a wider
exponential tail develops at the peak. This, by looking at the
difference of the peak and original distributions, arises from the
coalescence of original microcracks. By direct observation as such,
but also by looking at the distributions with the largest (or second
largest as well) removed it becomes clear that the tail is indeed
averaged over the largest cracks in each sample. 

\begin{figure}[!hbt]
\includegraphics{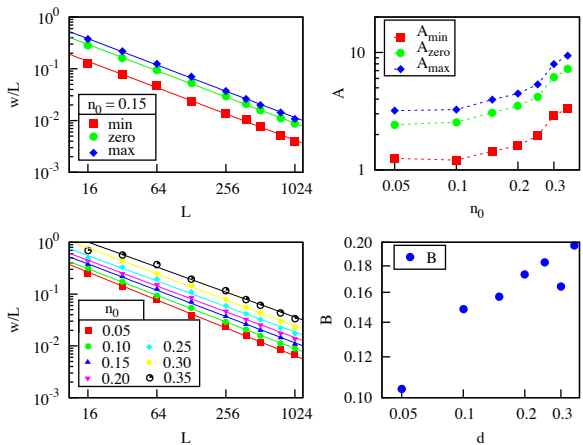}
\caption{(Color online) Mass transport of cracks upto the peak. Upper left: for
  $n_0=0.15$ the point of maximum and minimum of the difference showed
  in red in the inset of Fig.~\ref{fig:4} scale following a power law
  $w(L) = AL^B$ with the same slope. This is true for every disorder
  as showed in the lower left panel (just the maximum, here). In the
  right panels the parameters resulting from the fitting: the
  prefactor (up) and the exponent (down) increase with the disorder.}
\label{fig:5}
\end{figure}

Interestingly Figure \ref{fig:5} shows that the mass transport of
cracks follows for all $n_r$ a power-law scaling with $L$. The Inset
of Fig.~\ref{fig:4} allows to identify three particular values for
each $L$ and $n_r$: the negative 
minimum, the positive maximum, and where the difference of the
distributions is zero. They all three follow for each $n_r$ a
power-law scaling with the same exponent. The prefactors and power-law
exponents follow monotonic trends with disorder. 

\vspace{1cm}

\begin{figure}[!h]
\includegraphics{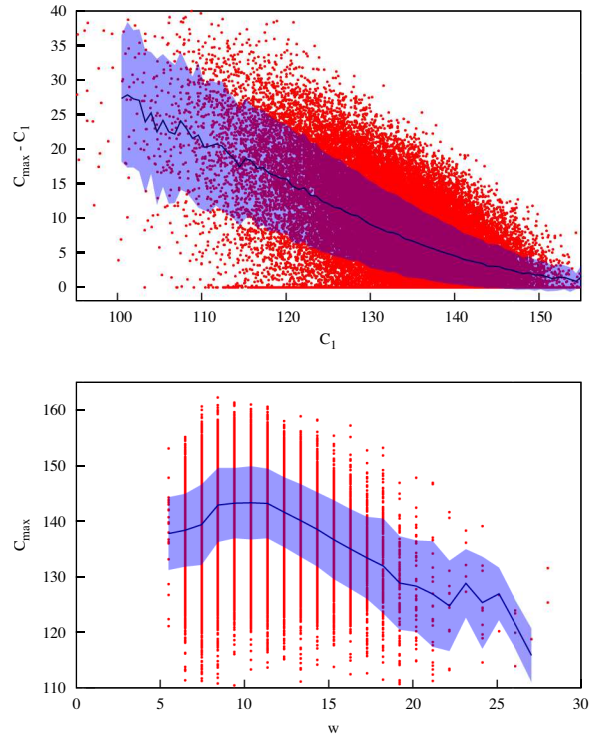}
\caption{(Color online) A scatter plot of peak strength values $C$ vs. the sample
  defect size $w$. The figure includes a 50-point running average over
  the $w$-values. The data represented here are obteined from $N=10^5$
  realization of a sample of size $L=512$ with initial disorder $n_r =
  0.10$.} 
\label{fig:6}
\end{figure}

Given that one can find the "largest crack" in each system, the
question remains how $w$ and $N_C$ correlate with the strength ($C$)
of each sample and with each other. Obviously there might be a slight
correlation between the two geometrical quantities, since the largest
cracks should develop if the system undergoes more failures $N$. The
initial step, with $N=1$ is related to the initial strength $C_1$ (or
$\sigma_1$ at which the first bond fails. 
There seems to be a slight anti-correlation in that a large $C_1$
implies a smaller final $w$. If the first bond breaks late, then
relatively speaking there is less crack growth. Figure \ref{fig:6}
shows the correlation of $C$ and $w$: obviously this does not follow
the Linear Elastic Fracture Mechanics prediction one-to-one, and it
moreover illustrates that the damage that develops is able to
decorrelate $C$ from $C_1$. 
This is not a novelty, since it is known that in the RFM $C_1$ follows
a similar modified Gumbel scaling \cite{Duxbury_et:1987} as $C$ but with
different parameters, and that there is no one-to-one correspondence
between $C$ and $C_1$, which would otherwise render fracture
prediction rather trivial \cite{Duxbury_et:1987}. However, the comparison of
$w$ with $C$ and $C_1$ allows to conclude that such a decorrelation
must be due to the fact that the final, critical defect experiences a
stress which has a random component.

\subsection{Crack coalescence}
The results presented above indicate that small amounts of damage is
enough to have a profound influence on the kind of dominating
microcracks at peak and thus also on the sample
strength. Theoretically, the question could be formulated by a
Smoluschowsky-like system of rate equations for defects of size $w$ to
a degree of some generality indeed. As we have already indicated, in
the current case --- dilution-like random disorder --- the most
important mechanism seems to be crack coalescence of fusion.

\begin{figure}[!h]
\includegraphics{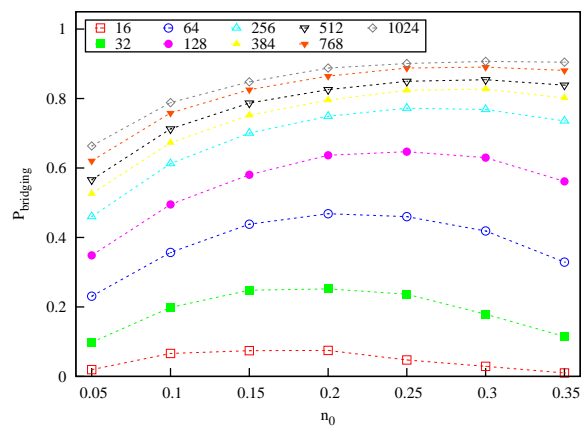}
\caption{(Color online) Bridging probability as a function of the
  disorder. The probability $P_{bridging}$ that a broken bonds joins
  two pre-existing cracks increases with the system size and, while is
  not clear the behaviour for the large systems, it presents a maximum
  at a given disorder value.}
\label{fig:7}
\end{figure}

The defining rates or processes for $P_{N+1,L} (l)$, when $N
\rightarrow N+1$ --- in other words a bond is broken --- are i)
joining two cracks, ii) crack growth ("$l\rightarrow l+1$"), and iii)
nucleation of a crack of size one ($l=1$). One can now check what the
effect of $N_C$ is like, and Figure \ref{fig:5} shows the likelihood
of the first of these three processes. Three major features emerge:
first, the microfailures have a large probability to contribute to
crack coalescence. Second, this increases with $L$. Third, there
appears again to be a maximum "efficiency", at a certain disorder,
which shifts slowly with $L$. Attempts to find a scaling form for
$P_{bridging} $ with $n_r$ and $L$ were not successful, leaving an
important unanswered question: does it saturate for a given disorder
to a value smaller than unity?

\section{Conclusions}

Here we have done the first systematic analysis of damage mechanics in
brittle fracture, using the Random Fuse Model as the theoretical
testbed. The approach is interesting for a number of reasons,
including the fact that the problem is extremely hard to study
experimentally to even a limited extent.  What transpires from our
results is that the peak-damage and the microcrack populations in our
samples at peak stress are determined by the microscopic dynamics of
crack growth, which occurs mostly by coalescence. This dynamics is
quantitatively dependent to a large degree on the disorder strength
and the system size. While the general question of size-effects and
fracture strength in these test systems is now finally
well-understood, the microscopic details here turn out to be quite
important: it is from them that the parameters of the coarse-grained
(modified) Gumbel distributions ensue. Thus the damage study at hand
here highlights a connection between the microscopic and macroscopic,
beyond  even that link which ensues during a renormalization or
upscaling of sample-size dependent strength distributions. 

A number of open avenues for future research must be listed. The
detailed connection of the damage scaling(s) and those of strength
distributions should be analyzed. The particular case at hand is
characterized by a very small damage degree at peak load. Scenarios
where the crack population undergoes more a complicated development
upto $C_{\textrm{max}}$ would be of great interest for further studies, such as
where the initial one has a power-law form. In a more general sense,
our results highlight the old engineering ideas of improving fracture
resistance by inhibiting crack growth: minute effects in the fracture
resistance may influence the strength quantitatively.

\section*{Acknowledgments}
S.Z. is supported by the European Research Council Advanced Grant 2011
- SIZEFFECTS and thanks the visiting professor program of Aalto
University. We acknowledge support and assistance by CSC — IT
Center for Science Ltd, Finland and generous resources and sponsorship
via the HPC Europe 2 and DEISA Consortium (EU pro jects FP6 RI-031513
and FP7 RI-222919). The Center of Excellence -program (COMP, Center of
Excellence in Computational Nanoscience) of the Academy of Finland is
thanked. We thank J. P. Sethna and A. Shekhawat for useful
discussions.

\bibliography{fracture_new}

\end{document}